\documentclass[12pt]{article}

\usepackage{amsmath}
\usepackage{graphicx,psfrag,epsf}
\usepackage{enumerate}
\usepackage{natbib}
\usepackage{url} % not crucial - just used below for the URL
\usepackage{kotex}
\usepackage{multirow}
\usepackage[small]{caption}
\usepackage{rotating}

\usepackage{color}
\usepackage{amssymb}
\usepackage{rotating}
\usepackage[multiple]{footmisc}

\usepackage[linesnumbered,ruled]{algorithm2e}

\newcommand{\blind}{0}

\newtheorem{assumption}{Assumption}

\usepackage{comment}
\usepackage{ulem}
\usepackage{soul}
\usepackage[english]{babel}
\usepackage{booktabs,longtable}
\definecolor{dark-red}{rgb}{0.4,0.15,0.15}
\definecolor{dark-blue}{rgb}{0.15,0.15,0.4}
\definecolor{medium-blue}{rgb}{0,0,0.5}

\def\bfD{{\ensuremath{\bf D}}}

\def\bfX{{\ensuremath{\bf X}}}

\def\bftheta{{\ensuremath\boldsymbol{\theta}}}

\def\bfalpha{{\ensuremath\boldsymbol{\alpha}}}
\def\bfbeta{{\ensuremath\boldsymbol{\beta}}}

\def\bfX{{\ensuremath{\bf X}}}

\def\bfalpha{{\ensuremath\boldsymbol{\alpha}}}
\def\bfbeta{{\ensuremath\boldsymbol{\beta}}}

\def\bfeta{{\ensuremath\boldsymbol{\eta}}}

\def\bftheta{{\ensuremath\boldsymbol{\theta}}}

\begin{document}

\def\spacingset#1{\renewcommand{\baselinestretch}%
	{#1}\small\normalsize} \spacingset{1}

%%%%%%%%%%%%%%%%%%%%%%%%%%%%%%%%%%%%%%%%%%%%%%%%%%%%%%%%%%%%%%%%%%%%%%%%%%%%%%

\if0\blind
{
	\title{\bf Estimating the effect of PEG in ALS patients using observational data subject to censoring by death and missing outcomes}
	\author{Pallavi Mishra-Kalyani\\
		Department of Biostatistics and Bioinformatics\\
		Emory University		\vspace{0.05in}\\
		Brent A. Johnson\\
		Department of Biostatistics and Computational Biology\\
		University of Rochester		\vspace{0.05in}	\\
	Jonathan D. Glass\\
		Department of Neurology\\
		Emory University		\vspace{0.05in}	\\
		and \\
		Qi Long\\
		Department of Biostatistics, Epidemiology, and Informatics\\
		University of Pennsylvania}
	\date{}
	\maketitle
} \fi

\if1\blind
{
	\bigskip
	\bigskip
	\bigskip
	\begin{center}
		{\LARGE\bf Estimating the effect of PEG in ALS patients using observational data subject to censoring by death and missing outcomes}
	\end{center}
	\medskip
} \fi

\bigskip
\begin{abstract}
	Though they may offer valuable patient and disease information that is impossible to study in a randomized trial, clinical disease registries also require special care and attention in causal inference. Registry data may be incomplete, inconsistent, and subject to confounding. In this paper we aim to address several analytical issues in estimating treatment effects that plague clinical registries such as the Emory  amyotrophic lateral sclerosis (ALS) Clinic Registry. When attempting to assess the effect of a surgical insertion of a percutaneous endoscopic gastrostomy (PEG) tube on body mass index (BMI) using the data from the ALS Clinic Registry, one must combat issues of confounding, censoring by death, and missing outcome data that have not been addressed in previous studies of PEG. We propose a causal inference framework for estimating the survivor average causal effect (SACE) of PEG, which incorporates a model for generalized propensity scores to correct for confounding by pre-treatment variables, a model for principal stratification to account for censoring by death, and a model for the missing data mechanism. Applying the proposed framework to the ALS Clinic Registry Data, our analysis shows that PEG has a positive SACE on BMI at month 18 post-baseline; our results likely offer more definitive answers regarding the effect of PEG than previous studies of PEG.
\end{abstract}

\noindent%
\vfill

\newpage
\spacingset{1.45} % DON'T change the spacing!

\section{Introduction}
\label{t3sec:intro}
\indent
While observational data from disease registries can be used to assess treatment effects, they are often fraught with analytical challenges. This is particularly true of a disease with high disability and mortality rates as issues of unmeasured data often occur due to a patient's absence or death. Our work is motivated by the analysis of data from an amyotrophic lateral sclerosis (ALS) Clinic Registry, aimed at estimating the effect of a palliative treatment, percutaneous endoscopic gastrostomy (PEG) on body mass index (BMI), a measure of nutritional management and quality of life. ALS is a neurodegenerative disorder with a very poor prognosis \citep{gelinas2000, procaccini2008}. Currently there is no known cure for ALS and existing treatment procedures for ALS patients are mostly palliative; there has been substantial interest in assessing effects of these palliative procedures on quality of life \citep{miller2009practice}.

Dysphagia, or difficulty in swallowing, affects almost all patients with ALS, and subsequently along with muscle atrophy and hypermetabolism causes malnutrition amongst the ALS patient population. Nutritional management is key in disease management and palliative care \citep{desport2006,muscaritoli2012}. PEG, a surgically inserted tube providing enteral nutrition, is generally considered when an ALS patient's nutritional status deteriorates and weight loss is greater than 10\% of the baseline weight \citep{goyal2014, park1992}, though patients may elect not to receive PEG for other reasons. While some studies have shown that PEG tube insertion offers some benefits in stabilizing or even increasing body weight, the overall results regarding the effect of PEG on weight or BMI are inconclusive, and in particular the long term benefits of the procedure are uncertain \citep{kasarskis1996,katzberg2011,mazzini1995}. One challenge for assessing the effect of PEG in ALS patients is that randomized studies are not ethically feasible, since PEG placement has long been recommended as standard of care in ALS clinics for patients with nutritional compromise \citep{miller2009practice}. As a result, the effect of PEG has been evaluated using observational data, and previous studies did not adequately address analytical issues including confounding and missing data that were commonly encountered. In addition, the existing studies had small to moderate sample sizes, limiting the generalizability of their findings.

The Emory ALS Clinic Registry has several notable strengths, making it well-suited for assessing the effect of PEG. It includes more than 1000 patients with long-term follow-up and a wide range of relevant clinical variables such as BMI, vital capacity (VC), forced vital capacity (FVC), negative inspiratory force (NIF), and ALS Functional Rating Scale Revised score (ALSFRS-R).  In particular, ALSFRS-R is a measure of disease progression and is not always collected in large ALS registries. However, the data present several analytical challenges to estimating a causal effect of PEG. First, insertion of PEG is not randomized, which may lead to confounding by pre-treatment variables. Second, due to the fatal and fast-progressing nature of ALS outcome measurements may be ``censored by death," i.e, a patient dies before the outcome of interest can be measured \citep{zhangandrubin2003, rubin2006causal}. Third, for those who are not ``censored by death,"   the outcome may be missing due to various reasons.

In the context of Rubin's Causal Model, principal stratification, first described in detail in \citet{frangakisandrubin2002}, offers a framework for causal inference in the presence of a confounding post-treatment variable, such as censoring by death. Subsequently, \citet{zhangandrubin2003} extended this approach to cases where a post-treatment variable such as survival or graduation ``censors" the outcome of interest. \citet{zhangetal2009} further outlined specific modeling approaches for the identification of survivor average causal effect (SACE) within the principal stratification framework. These existing works in this area, however, analyzed data from randomized studies, assuring complete collection of outcome data and, more importantly, avoiding the issues of confounding, an advantage not guaranteed in our retrospective clinic registry data.

More recently, \citet{frumento2012} addressed post-treatment variables and missing outcome data simultaneously within a principal stratification framework in a randomized study. However, the Emory ALS Clinic Registry data introduce some unique challenges that were not present in the analysis of \citet{frumento2012} and require new analytical approaches. In particular, our study treatment is not randomized and includes a component of time to treatment, leading to potential confounding by pre-treatment variables. Furthermore, both treatment and outcome measurement are subject to censoring by death in our study. Finally, missing outcome data in our study are due to the observational nature of the study, and have different implications compared to the missing outcomes in \citet{frumento2012}. Therefore, while some of the same tools are utilized in our work, the proposed framework addresses important new issues.

It is particularly important to address potential confounding by pre-treatment variables in a principal stratification model, because \citet{frangakisandrubin2002} defined principal stratification in a randomized experiment. When such confounding may be present, either as residual confounding in a randomized clinical trial or due to observational data, \citet{schwartz2012} showed that the resulting principal effect estimate is likely to be biased. Of the many methods for handling confounding, a popular choice is the propensity score introduced by \citet{rosenbaumandrubin1983}. The propensity score provides a means of balancing pre-treatment variables across treatment groups, a result that would otherwise be guaranteed if a randomized study design is used. Though propensity scores were initially introduced for balancing treatment assignment groups when treatment is binary, other authors have extended these methods to non-binary treatment assignment models, such as generalized propensity score methods \citep{imaiandvandyk2004,hirano2004}. These methods allow for balancing pre-treatment variables when treatment assignment is categorical or continuous. A recent work by \citet{jo2009} used propensity scores within a principal stratification framework, not for the removal of confounding as the study was randomized, but instead to predict principal strata membership in a matched analysis, again different from our setting. Furthermore, propensity score methods for non-binary treatment assignment have not yet been employed for conditional ignorability in a principal stratification framework when outcomes may be missing.

To account for aforementioned complicating analytical issues, we develop a causal inference approach for estimating the survivor average causal effect (SACE) of PEG. Building on the principal stratification framework to account for censoring by death (i.e., confounding by a post-treatment variable), our approach uses generalized propensity scores to correct for confounding by pre-treatment variables and it also accounts for missing outcome data by incorporating a model for the missing data mechanism. Although it is developed for the specific application, our approach can be applied to analysis of other observational studies that are subject to similar issues. The remainder of the article is organized as follows. First, we introduce the data and notation in Section~\ref{t3sec:notation}. We present our framework for causal inference in Section~\ref{t3sec:methods} and a Bayesian inference procedure in Section~\ref{t3sec:bayes}. In Section~\ref{t3sec:app}, we apply the proposed approach to the Emory ALS data. We conclude this paper with some discussion remarks in Section~\ref{sec:dicussion}.

\section{Data Structure in the ALS Registry}
\label{t3sec:notation}

In the analysis of the ALS Clinic Registry, our goal is to assess the effect of PEG on BMI (denoted by $Y$) at $t^o=18$ months post-baseline; without loss of generality we assume that $Y\in \mathcal{R}$. We denote by $\bfD$ the set of patient baseline characteristics and by $T_S$ the survival time post-baseline confirmed by Social Security database records, noting that all patients in our data set died. The surgical insertion of a PEG tube may be administered at any time post-baseline given patient choice. Let $Z$ be the binary indicator such that $Z=1$ if a patient received PEG prior to the time of outcome measurement $t^o$ and $Z=0$ if otherwise; let $T_Z$ denote time to treatment. It follows that patients untreated prior to $t^o$ have values $Z=0$ and $T_Z=T_S$ if they did not survive until $t^o$, or $Z=0$ and $T_Z = t^o$ if they survived past $t^o$. For those who died prior to $t^o$, $Y$ is undefined and is known as ``censored by death." For those who are alive at $t^o$, $Y$ is defined but may be missing due to other reasons and let $M$ be the binary indicator such that $M=1$ if $Y$ is missing  and $M=0$ if $Y$ is observed. Of note, for patients with $S=0$ who were ``censored by death," both $Y$ and $M$ are undefined, denoted by ``*", following the notation of \citet{zhangandrubin2003}. This extends the sample space for $Y$ to \{$\mathcal{R},*$\} and for $M$ to \{$0,1,*$\}.

\section{Causal Inference Framework}
\label{t3sec:methods}
%\subsection{Assumptions \qld{and Causal Estimand}}

As earlier defined, the binary indicator for treatment from baseline until $t^o$ (time of outcome measurement) is $Z$ and the outcome of interest is $Y$. A post-treatment indicator of survival past the $t^o$ is $S$. Following the Rubin Causal Model \citep{holland1986}, we define potential outcomes $\mathcal{Y} = \{\left(Y_i(0),Y_i(1)\right) \mbox{ for } i = 1 \ldots n\}$ and $\mathcal{S} = \{\left(S_i(0), S_i(1)\right)\mbox{ for } i = 1 . . . n\}$ under the two treatment options $Z=0$ and 1. For simplicity, we suppress $i$ in what follows if it does not create ambiguity.

\begin{figure}
%\centering
\includegraphics[height=7.8cm]{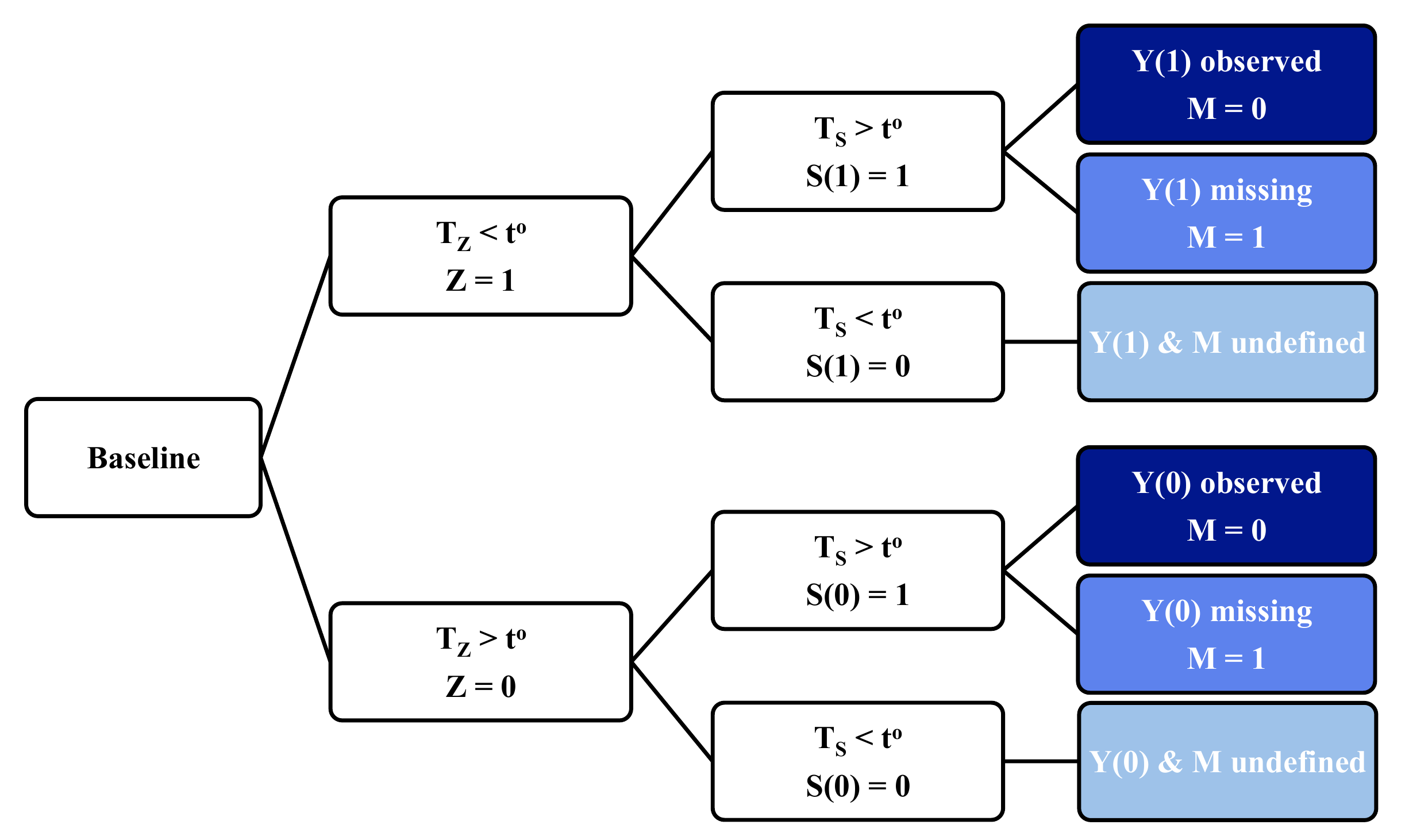}
\caption{\label{fig:missing} Potential outcomes for all patients given survival and treatment status.}
\end{figure}

Figure \ref{fig:missing} illustrates the potential outcomes in the Emory ALS data. For example, patients who are treated prior to time of outcome measurement ($T_Z > t^o$) may follow one of two paths: survival until $t^o$ ($S$ = 1) or death prior to $t ^o$ ($S$ = 0). Those individuals who survive may have observed $Y$, as designated by the darkest shaded box in Figure \ref{fig:missing}, or have missing $Y$ as designated in the medium shaded box. Patients who die before $t^o$ have both $Y$ and $M$ undefined, i.e., censored by death, as designated by the lightest shaded box.

In the presence of censoring by death, we use post treatment survival status to define four potential principal strata ($G$) by $\left(S_i(0), S_i(1)\right)$, namely, $LL$, $LD$, $DL$, and $DD$. Individuals who are in the $LL$ stratum are those who would be alive at time $t^o$ regardless of treatment. Those individuals who are in the $LD$ stratum would be alive if they received the treatment but would not be alive if they did not and those who are in the $DL$ stratum would not be alive if they receive the treatment, but would be alive if they did not. Finally, individuals in the $DD$ are those who would die prior to time $t^o$ regardless of treatment. A causal effect of treatment is only estimable in strata where individuals would be alive and therefore have a measurable outcome, regardless of treatment status. As such, the causal effect is estimable in the $LL$ stratum only. Specifically, the Survivor Average Causal Effect (SACE) is defined as the difference in $Y$ between treated and untreated individuals in the $LL$ stratum.

However, we do not get to observe $G$ since we only know the survival status given the observed treatment status. Based on $(Z,S)$, there are four observed groups denoted by $O(Z,S)$: $O(1,1)$, $O(1,0)$, $O(0,1)$, and $O(0,0)$. Each of these observed groups is a mixture of the principal strata. Specifically, $O$(1,1) is a mixture of individuals from the $LL$ and $LD$ strata, $O$(1,0) is comprised of individuals from the $LL$ and $DL$ strata, $O$(0,1) is comprised of individuals from the $DD$ and $DL$ strata, and $O$(0,0) is comprised of individuals from the $DD$ and $LD$ strata. In addition, based on $(Z,S,M)$, we have six observed groups denoted by $O$($Z,S,M$), namely, $O$(1,1,1), $O$(1,1,0), $O$(1,0,-), $O$(0,1,1), $O$(0,1,0), $O$(0,0,-); see Table \ref{t3tab:like}. The two original observed groups with $S=1$, namely, $O$(1,1) and $O$(0,1), are each divided into two subgroups based on whether $Y$ is observed or missing. The groups with $S =0$ do not need to be divided further, since $M$ is undefined.

\subsection{Assumptions}
In our causal inference framework, we adopt two standard assumptions, the Stable Unit Treatment Value Assumption (SUTVA) as defined by Cox in 1958 and further elaborated in \citet{rubin1980} and the Strong Ignorability of Treatment Assignment \citep{rosenbaumandrubin1983}.
\begin{assumption}\label{A1} [SUTVA] The potential outcomes of any unit is independent of the treatment status of the other units given $\bfD$.
\end{assumption}
\begin{assumption}\label{A2} [Strong Ignorability of Treatment Assignment]  $\left(Y_i(0),Y_i(1)\right)$ and $\left(S_i(0), S_i(1)\right)$ (or equivalently $G$) are independent of $Z$ given $\bfD$.
\end{assumption}

Regarding the missing data mechanism for $Y$, we consider two assumptions, a latent ignorability \citep{frangakisandrubin1999} and a standard ignorability \citep{Little02}.
\begin{assumption}\label{A3} [Latent Ignorable Missingness]  $M$ does not depend on the missing values or parameters of the data distribution conditional on $\bfD$ and $G$; particularly, $M$ is independent of $Y$ given $\bfD$ and $G$.
\end{assumption}
\begin{assumption}\label{A4} [Ignorable Missingness]  $M$ does not depend on the missing values or parameters of the data distribution conditional on $\bfD$; particularly, $M$ is independent of $Y$ given $\bfD$.
\end{assumption}
Since $G$ is not completely known, the latent ignorable missingness in Assumption 3 is not ignorable and it is necessary to model $M$ in the modeling framework.

\subsection{Data Likelihood}
We define the complete data as $(Y_i,S_i,M_i,G_i,Z_i,\bfD_i)$ for $i=1,\ldots,n$, which under Assumption~\ref{A1} constitute independent and identically distributed observations. Accordingly, the observed data are represented as $(Y_i,S_i,M_i,G_{obs,i},Z_i,\bfD_i)$, where $G_{obs,i}$ is the set of feasible principal strata that an individual may belong to given $Z_i$ and $S_i$.

We first relate the distribution of the complete data $(Y_i,S_i,M_i,G_i,Z_i,\bfD_i)$  to the distribution of the potential outcomes under Assumptions 1-3. We have
\[f(Y_i|S_i,M_i,G_i,Z_i,\bfD_i)=f(Y_i|M_i,G_i,Z_i,\bfD_i)=f(Y_i|G_i,Z_i,\bfD_i)=f(Y_i(Z_i)|G_i,\bfD_i)\]
where the first equality is due to the fact that $S_i$ is redundant given $Z_i$ and $G_i$, the second equality is due to Assumption~\ref{A3}, and the third equality is due to Assumption~\ref{A2}. In addition, $f(S_i|M_i,G_i,Z_i,\bfD_i)=1$, since $S_i$ is fully determined by $Z_i$ and $G_i$. From Assumption~\ref{A2}, we have $f(G_i|Z_i,\bfD_i)=f(G_i|\bfD_i)$. Using these facts and Assumption~\ref{A1}, the complete data likelihood can be written as
\begin{eqnarray}
&&\prod_i f(Y_i,S_i,M_i,G_i,Z_i|\bfD_i)\nonumber\\
&=&\prod_i f(Y_i|S_i,M_i,G_i,Z_i,\bfD_i)f(S_i|M_i,G_i,Z_i,\bfD_i)f(M_i|G_i,Z_i,\bfD_i)f(G_i|Z_i,\bfD_i)f(Z_i|\bfD_i)\nonumber\\
&=&\prod_i f(Y_i(Z_i)|G_i,\bfD_i)f(M_i|G_i,Z_i,\bfD_i)f(G_i|\bfD_i)f(Z_i|\bfD_i),\label{Complete-Data1}
\end{eqnarray}
where $f(Y_i(Z_i)|G_i,\bfD_i)$ models the potential outcomes, $f(M_i|G_i,\bfD_i)$ models the missing data mechanism, $f(G_i|\bfD_i)$ models the principal strata, and $f(Z_i|\bfD_i)$ models the treatment assignment. Without loss of clarity, we define $f(Y_i|G_i,\bfD_i)=1$ if $Y_i$ is missing; in other words, only observed $Y_i$ contributes to the first part of the complete data likelihood \eqref{Complete-Data1}. Assuming the parameters in $f(Z_i|\bfD_i)$ are distinct from the other parameters and $f(Z_i=z|\bfD_i)>0$ for $z=0,1$, we can drop $f(Z_i|\bfD_i)$ from the complete data likelihood.

The observed data likelihood is considerably more complicated, since it involves mixture distributions since we only observe $G_{obs,i}$ and do not observe $G_{i}$. As shown in Section~\ref{subsec:inference}, a data augmentation algorithm allows us to work with the complete data likelihood directly and hence avoid the complicated observed data likelihood.

%\subsection{Generalized Propensity Scores}
Since the treatment of interest includes a time-to-intervention component $T_Z$, we use generalized propensity scores in the spirit of \citet{imaiandvandyk2004}, \citet{lu2005} and \citet{hirano2004} to further accommodate this unique feature. Specifically, we model $T_Z$ which may be censored by $T_S$ using a Cox proportional hazards (PH) model $h(t) = h_0(t)\exp\left(\bfbeta^T\bfD\right)$ and use its linear predictor $\bfbeta^T\bfD$ as the generalized propensity score, denoted by $PS$. Along the lines of \citet{imaiandvandyk2004} and \citet{lu2005}, it follows from Assumptions~\ref{A1} and~\ref{A2} that $Z$ is independent of $\left(Y_i(0),Y_i(1)\right)$ and of $\left(S_i(0), S_i(1)\right)$ conditional on $PS_i$. Then the complete data likelihood~\eqref{Complete-Data1} can be rewritten as
\begin{eqnarray}
\prod_i f(Y_i(Z_i)|G_i,PS_i)f(M_i|G_i,Z_i,\bfD_i)f(G_i|PS_i),\label{Complete-Data2}
\end{eqnarray}
where $f(Z_i|\bfD_i)$ is dropped.

Alternatively, under Assumptions 1, 2 and 4, i.e., replacing the latent ignorable missingness with the ignorable missingness, the complete data likelihood~\eqref{Complete-Data2} can be simplified to \[\prod_i f(Y_i(Z_i)|G_i,PS_i)f(M_i|Z_i,\bfD_i)f(G_i|PS_i).\]
Assuming that its parameters are distinct from parameters in other models, $f(M_i|Z_i,\bfD_i)$ can be dropped from the complete data likelihood, leading to
\begin{eqnarray}
\prod_i f(Y_i(Z_i)|G_i,PS_i)f(G_i|PS_i).\label{Complete-Data3}
\end{eqnarray}
In Section 5, we analyze the ALS data using both \eqref{Complete-Data2} and \eqref{Complete-Data3} under the latent ignorable missingness and the ignorable missingness, respectively.

\section{Modeling Framework}
\label{t3sec:bayes}
In this section, we specify the model for each component of the complete data likelihood~\eqref{Complete-Data2} and then describe a Bayesian inference approach under Assumptions 1-3. The corresponding inference approach under Assumptions 1-2 and 4 is simpler and can be derived along similar lines.

\subsection{Model for Potential Outcomes}
\label{sec:outcome}
For $f(Y_i(Z_i)|G_i,PS)$ in \eqref{Complete-Data2}, we assume that $\left(Y(1),Y(2)\right)$ within each principal stratum $G=g$ have a normal distribution, $f_{g}$, with parameters $\bfeta_{g}$ and covariates $\bfX_{1,g}$ that may differ by strata. Also, individual likelihoods for $Y$ are dependent on potential strata $g$, and are represented by $f_{g,i}$. Specifically, $Y \sim N(\bfX_{1,g}\bfeta_{g},\sigma_{g}^2)$ for $g \in {LL, LD, DL}$, where $\bfX_{1,LL}$ includes the column for intercept, $Z$, $T_Z$, and $PS$. $\bfX_{1,LD}$ and $\bfX_{1,DL}$ include intercept and $PS$ but not $Z$ and $T_Z$ as patients with an observed outcome in each of these strata are either all treated or all untreated respectively. The regression coefficient for $Z$ in the $LL$ stratum represents SACE and is of primary interest. Inclusion of $T_Z$ allows us to assess the relationship between $Y$ and time to treatment within those who receive the treatment.

\subsection{Model for Principal Stratification}
\label{t3sec:strata}
The probabilities of the four principal strata, denoted by $(\pi_{LL}, \pi_{LD}, \pi_{DL}, \pi_{DD})$, are modeled using a multinomial logit model with a set of covariates denoted by $\bf{X}_2$ including $PS$. he probability of a patient being in principal stratum $g$ is given in equation (\ref{t3eq:probg}), and individual probabilities of principal strata are represented by $\pi_{g,i}$. As in any multinomial logit model, one category must be selected as a reference group.
\begin{equation}
\label{t3eq:probg}
\pi_{g} = P(G = g|\bfX_{2}) = \frac{\exp(\bfX_{2}^T\bfalpha_g)}{\sum\nolimits_{g'}\exp\left(\bfX_{2}^T\bfalpha_{g'}\right)}
\end{equation}
Of the assumptions available for sharpening the bounds of effect estimates in the principal stratification framework, a monotonicity assumption is often used, i.e., one of the principal strata, say the $DL$ group does not exist \citet{zhangandrubin2003}. While this monotonicity assumption may be plausible if treatment is always beneficial on patient survival, it is also possible that a subpopulation of weak patients would be not be able to withstand treatment or recovery and may die shortly after treatment, but could stay alive if left untreated, when evaluating an intrusive treatment. For such a data analysis, it may be appropriate to allow all four principal strata to exist and to relax the assumption of monotonicity.  In the ALS data, the SACE for PEG is estimated both with and without the monotonicity assumption to assess its impact.

\subsection{Model for Missing Data Mechanism}
\label{sec:missing}
%Thus far, our framework assumes that all those individuals alive at $t^o$ will have an observed outcome. In studies such as clinical trials or other prospective study designs, it may be appropriate to expect outcome measurements at planned time-points. However, the reality of retrospective studies is that often measurements of patient status are available at sporadic times, which may not align with the research question of interest. Straightforward options for defining the outcome of interest may either be too broad, by including any outcome observed from baseline until the time of interest, or may be too exclusive, by only including those outcomes observed in a narrow window around the time of interest. This latter definition of the outcome is specific and does allow for a precise analysis of the effect of treatment on outcome, however by excluding those individuals who do not have an outcome in the time period of interest from the analysis, we squander the information that these individuals may provide. Instead, the outcome is defined for those individuals who do not have a measurement during the time frame of interest as missing.

As discussed before, $M$ for patients with $S=0$ is undefined, so it is only valid to model $M$ for those individuals in the $LL$ stratum with treatment, the $LL$ stratum without treatment, the $LD$ stratum with treatment, and the $DL$ stratum without treatment. The model for $M$ is presented in \eqref{eq:probm}.
\begin{equation}
\label{eq:probm}
\phi_{gz} = Pr(M = 1|G=g, Z=z, S=1) = \frac{e^{\bfX_{3}\bftheta_{g,z}}}{1+e^{\bfX_{3}\bftheta_{g,z}}}
\end{equation}
where $\bfX_{3}$ is a set of covariates associated with the missing data mechanism.  The individual probability of missingness is dependent on strata and treatment value and is represented by $\phi_{gz,i}$. When Assumption 4 is used instead of Assumption 3, this model is not needed.

\begin{table}[h]
\centering
\caption{Individual likelihood by observed groups if $G_i$ is known}
\begin{tabular}{ccccc}
\hline
& \multicolumn{4}{c}{$G_i$} \\
\hline
Observed  & & & & \\
Group $O(Z,S,M)$ &$LL$ & $LD$ & $DL$ & $DD$ \\
\hline
\hline
\multirow{2}{*}{$O(1,1,0)$} & $(1-\phi_{LL,1,i})\times$ & $(1-\phi_{LD,1,i})\times$ & \multirow{2}{*}{-} & \multirow{2}{*}{-} \\
 & $\pi_{LL,i}f_{LL,i}$ & $\pi_{LD,i}f_{LD,i}$ &  &  \\
\multirow{2}{*}{$O(1,1,1)$} & \multirow{2}{*}{$\phi_{LL,1,i}\pi_{LL,i}$} & \multirow{2}{*}{$\phi_{LD,1,i}\pi_{LD,i}$} & \multirow{2}{*}{-} & \multirow{2}{*}{-} \\
 &  &  &  &  \\
\multirow{2}{*}{$O(1,0,-)$} & \multirow{2}{*}{-}  & \multirow{2}{*}{-}  & \multirow{2}{*}{$\pi_{DL,i}$} & \multirow{2}{*}{$\pi_{DD,i}$} \\
 &  &  &  &  \\
\multirow{2}{*}{$O(0,1,0)$} & $(1-\phi_{LL,0,i})\times$ & \multirow{2}{*}{-}  & $(1-\phi_{DL,0,i})\times$ & \multirow{2}{*}{-}  \\
 & $\pi_{LL,i}f_{LL,i}$ &  & $\pi_{DL,i}f_{DL,i}$ & \\
\multirow{2}{*}{$O(0,1,1)$} & \multirow{2}{*}{$\phi_{LL,0,i}\pi_{LL,i}$} & \multirow{2}{*}{-}  & \multirow{2}{*}{$\phi_{DL,0,i}\pi_{DL,i}$} & \multirow{2}{*}{-}  \\
 &  &  &  &  \\
\multirow{2}{*}{$O(0,0,-)$} & \multirow{2}{*}{-}  & \multirow{2}{*}{$\pi_{LD,i}$} & \multirow{2}{*}{-} & \multirow{2}{*}{$\pi_{DD,i}$} \\
 &  &  &  &  \\
\hline
\label{t3tab:like}
\end{tabular}
\end{table}

\subsection{Bayesian Inference}\label{subsec:inference}

Given the models specified in Section~\ref{sec:outcome}-\ref{sec:missing}, the individual likelihoods for all possible combinations of $Z$, $S$, and $M$ are given in Table~\ref{t3tab:like}, where each cell value is the likelihood of the complete data if $G$ is known. The conditional probability of $\mathnormal{G_i = g}$ given the observed data is the ratio of each cell to the total of that row. Rows $O(1,0,-)$ and $O(0,0,-)$ are included in Table~\ref{t3tab:like} for a comprehensive understanding of the possible combinations of treatment and survival, but individuals who fall into these groups do not have outcome data and hence do not contribute to the observed data likelihood. As such, these individuals will only contribute to the model for the probability of principal strata, which is reflected in the observed data likelihood in equation (\ref{t3eq:obsdatalikelihood}).
\begin{equation}
\label{t3eq:obsdatalikelihood}
\begin{aligned}
&\prod_{i \in O(1,1,0)} \left\{(1-\phi_{LL,1,i})\pi_{LL,i}N(\bfX_{1,LL,i}\eta_{LL},\sigma^2_{LL}) + (1-\phi_{LD,1,i})\pi_{LD,i}N(\bfX_{1,LD,i}\eta_{LD},\sigma^2_{LD})\right\}\times\\
&\prod_{i \in O(0,1,0)} \left\{(1-\phi_{LL,0,i})\pi_{LL,i}N(\bfX_{1,LL,i}\eta_{LL},\sigma^2_{LL}) + (1-\phi_{DL,0,i})\pi_{DL,i}N(\bfX_{1,DL,i}\eta_{DL},\sigma^2_{DL})\right\}\times\\
&\prod_{i \in O(1,1,1)} \left\{(\phi_{LL,1,i})\pi_{LL,i} + (\phi_{LD,1,i})\pi_{LD,i}\right\}\times\\ &\prod_{i \in O(0,1,1)} \left\{(\phi_{LL,0,i})\pi_{LL,i} + (\phi_{DL,0,i})\pi_{DL,i}\right\}\times\\
&\prod_{i \in O(1,0,-)} \left\{\pi_{DD,i} + \pi_{DL,i}\right\} \prod_{i \in O(0,0,-)}\left\{\pi_{DD,i} + \pi_{LD,i}\right\}
\end{aligned}
\end{equation}

Prior distributions for the specified parameters in the observed data likelihood should be chosen carefully, with thought to distributions that may be informative, proper, and conjugate where appropriate. For this analysis, a conjugate multivariate normal prior distribution is assumed for $\bfeta_g$, $\bfeta_g \sim Normal_p\left(\mu_g,\sigma_g^2V_g\right)$. Similarly, a conjugate prior distribution is assumed for $\sigma_g^2$, $\sigma_g^2 \sim Inverse Gamma\left(\nu_g,\omega_g\right)$. We choose diffused priors for $\bfalpha_g$ and $\bftheta_{g,z}$.
%$p\left(\eta_g\right) \propto |\sigma_g^2V_g|^{-\frac{1}{2}}e^{-\frac{1}{2\sigma_g^2}\left(\eta_g-\mu_g\right)^TV_g^{-1}\left(\eta_g-\mu_g\right)}$
%$p\left(\sigma_g^2\right) \propto \sigma_g^{2\left(-\nu_{LL}-1\right)}exp\left(-\frac{\omega_{LL}}{\sigma_{LL}^2}\right)$

Though the principal stratum of each individual is unknown, the observed treatment and survival groups may be used to inform imputation of the principal stratum assignments. Applying this idea, we use the Data Augmentation (DA) algorithm \citep{tanner1987} for posterior computation, in which information about the latent groups, namely, $G$, is imputed and subsequently the posterior parameters distributions are simulated.

The DA algorithm includes two iterative and alternating steps to allow for posterior inference. The first step, the Imputation or I-step, imputes the value of the principal stratum $G$ for each individual. This is accomplished by using the parameter values $\bfalpha_g^{\left(k\right)}$, $\bfeta_g^{\left(k\right)}$, $\sigma_g^{2\left(k\right)}$, and $\bftheta_{g,z}^{\left(k\right)}$ (for approach 2 only) from the current approximation of posterior (from the $k$th iteration) to generate $G^{\left(k+1\right)}$ by using the conditional probabilities that are given by taking the ratio of each cell value to the row total in Table~\ref{t3tab:like}. These conditional probabilities, $\rho_{O}$, are used in a Bernoulli distribution that imputes individual memberships to one of the two principal strata that correspond with the observed group $O$. More specifically, at the ($k$+1) iteration, each individual has a probability of being in a stratum that depends on their observed values ($Z$, $S$, $M$, $Y$, $PS$).

The P-step, or Posterior step, is then employed by using the imputed complete data set, and the parameters $\{\bftheta^{\left(k\right)}, \left(\pi_g^{\left(k\right)}, \bfeta_g^{\left(k\right)}, \sigma_g^{2\left(k\right)}\right)\}$ can be updated to $\{\bftheta^{\left(k+1\right)}, \left(\pi_g^{\left(k+1\right)}, \bfeta_g^{\left(k+1\right)}, \sigma_g^{2\left(k+1\right)}\right)\}$ by sampling from the full conditional distributions of each parameter based on the complete data likelihood. Either a Gibbs Sampler or a Metropolis-Hastings (MH) Algorithm can be employed for sampling. The details of the DA algorithm are provided in Web Appendices A and B.

If we use Assumption 4 instead of Assumption 3, we can drop all terms involving $\phi_{g,z}$ from Table~\ref{t3tab:like} and use the complete data likelihood \eqref{Complete-Data3}, leading to a simpler version of the aforementioned DA algorithm.

\section{Analysis of ALS Clinic Data}
\label{t3sec:app}

We apply the proposed approach to analysis of the data from patients who visited the Emory ALS clinic at least once between 1997 and 2014. Patients are excluded from the analysis for not having any follow up clinic visits from baseline to outcome measurement at month 18 post-baseline or for having long survival times ($>$5 years post-baseline), resulting in a cohort of $N=815$ for our analyses. It has been noted that ALS patients with long survival times are likely different from the other ALS patients \citep{mateen2010}, which may not be fully explained by clinical variables collected. Characteristics measured at baseline for each individual include sex, race, age, site of disease onset, months from symptom onset to baseline, months from diagnosis to baseline, body mass index (BMI), vital capacity (VC), forced vital capacity (FVC), negative inspiratory force (NIF), and ALS Functional Rating Scale Revised score (ALSFRS-R). Continuous variables, namely age, months from symptom onset to baseline, months from diagnosis to baseline BMI, VC, FVC, NIF, and ALSFRS-R, are standardized before being included as covariates for the propensity score model and the modesls for principal stratification model, missing data mechanism and outcomes.  Since our focus is to handle missing outcome values, we impute missing values in covariates prior to subsequent analysis using the proposed approach.
%These variables also have missing values, ranging from $<$1\% for sex to 60\% for ALSFRS-R score.

Of the 275 treated patients, 32\% or 89 individuals are alive 18 months from baseline, while of the 540 untreated individuals, 34\% or 186 individuals are alive at this time-point (p = 0.606). Baseline measurements of BMI, FVC, and VC, as well as age at diagnosis, survival until $t^o$, and proportion of white versus other races  are not significantly different between treated and untreated groups. However, clinically relevant differences between treatment groups do exist. NIF and ALSFRS-R scores at baseline are significantly higher for treated versus untreated individuals at a level of $\alpha = 0.05$. Also the proportions of female individuals and of individuals with spinal disease are significantly differnt for treated individuals than untreated individuals (p = 0.040 and p$<$ 0.001).  Most interestingly, the na\"ive analysis that compares the outcome of interest, BMI at 18 months post-baseline, across treatment groups indicates that treated individuals have a BMI that is 1 unit lower than the untreated group, and this difference in significant (p = 0.021). These results suggest that the effect of PEG may be confounded by variables that are different between treated and untreated groups, hence it is important to use the generalized propensity score. A table of patient characteristics by treatment status is available in the online Appendix (Web Table 1).

The 275 survivors at $t^o$ are significantly different from 540 non-surviving individuals by nearly every clinical comparison, with only demographic characteristics of sex and race being similar amongst the two groups, suggesting the need to adjust for the post-treatment variable, censoring by death. Additionally there are significant differences amongst patients with observed outcome and those with missing outcome data in, for example, BMI, FVC, VC, NIF, and ALS FRS-R score at baseline. indicating that the missing completely at random (MCAR) assumption is unrealistic and necessitates the use of the proposed approach for handling missing data. Web Tables 2 and 3 providing support for these statements are available in the online Appendix.

%\subsection{Estimation of SACE of PEG Treatment}
The SACE of PEG treatment is estimated using three modeling approaches. In addition to the two approaches using Assumptions 3 (latent ignorable missingness) and 4 (ignorable missingness) as described in Section \ref{sec:missing}, we consider a third approach that includes only individuals with an outcome measurement at $t^o$, analogous to a complete-case analysis that essentially assumes that the outcomes are MCAR. We also apply the same set of methods without the propensity score adjustment as well as with a monotonicity assumption, i.e., removal of the $DL$ stratum \citep{zhangandrubin2003}.  Finally, to allow for flexibility, polynomial splines are used for incorporating the generalized propensity score in the models with $d$ denoting the degree of the polynomial spline. For all analyses, the MCMC algorithm is run for a total of 5,000 iterations, with a burn-in period of 3000 iterations.

To identify the model that best fits the data, we use the deviance information criterion (DIC) \citep{spiegelhalter2002}. Models with smaller values of DIC are preferred to models with larger values, and the best fit is indicated by a minimum DIC. Table \ref{t3tab:dic} presents DIC for each of the modeling approaches under different assumptions for missing data mechanism with various degrees of polynomial splines for the propensity score. The models for all three approaches with $d=4$ have the lowest DIC, and therefore offer the best fit for each modelling approach. Particularly, the model assuming latent ignorable missingness with $d=4$ has the lowest DIC among all models.

\begin{table}[h]
  \centering
  \caption{\label{t3tab:dic} Deviance Information Criterion (DIC) for analysis under Assumption 3 (latent ignorable missingness), Assumption 4 (ignorable missingness) and MCAR, as $d$, the degree of polynomial splines for propensity scores, varies.}
    \begin{tabular}{rcccccc}
 \hline
  & \multicolumn{6}{c}{\textbf{\textit{d}}} \\
  &   \textit{0} &\textit{1} & \textit{2} & \textit{3} & \textit{4} & \textit{5}\\
 \hline
   Assumption 3 & 1702.55 & 1683.98 & 1668.20 & 1777.01 & 1658.00 & 2034.26 \\
   Assumption 4 & 1769.40 & 1722.59 & 1727.66 & 1733.74 & 1714.54 & 1719.51 \\
      MCAR &  2715.46 & 2658.06 & 2655.16 & 2654.74 & 2653.52 & 3200.87\\
 \hline
    \end{tabular}%
  \label{tab:addlabel}%
\end{table}

\begin{table}[h]
\centering
\caption{\label{t3tab:results} Estimates of SACE of PEG treatment (with 95\% credible intervals) on BMI at 18 months post-baseline Assumption 3 (latent ignorable missingness), Assumption 4 (ignorable missingness) and MCAR with and without adjustment for propensity scores (PS) (N=815).}
    \begin{tabular}{llccc}
    \hline
       & & \multicolumn{3}{c}{\textbf{\textit{Effect Estimates}}} \\
      & & \textit{Latent Ignorable}  & \textit{Ignorable} & \textit{} \\
      & & \textit{Missingness
}  & \textit{Missingness} & \textit{MCAR} \\
    \hline
    \multirow{4}[0]{*}{With PS} & \multirow{2}[0]{*}{Time to PEG} & 0.17 & 0.16 & 0.14 \\
       &    & \textit{(-0.06,0.39)} & \textit{(-0.05,0.38)} & \textit{(-0.01,0.29)} \\
\cline{2-5}
       & \multirow{2}[0]{*}{Binary Indicator} & 3.09 & 3.17 & 2.28 \\
       &    & \textit{(0.01,5.83)} & \textit{(0.14,6.24)} &\textit{(0.79,3.77)} \\
\hline
    \multirow{4}[0]{*}{Without PS} & \multirow{2}[0]{*}{Time to PEG} & 0.24 & 0.24 & 0.20 \\
       &    & \textit{(0.03,0.46)} & \textit{(0.03,0.45)} & \textit{(0.03,0.35)} \\
\cline{2-5}
       & \multirow{2}[0]{*}{Binary Indicator} & 0.45 & 0.56 & -0.05\\
       &    & \textit{(-2.42,3.45)} & \textit{(-2.19,3.31)} & \textit{(-1.37,1.23)}\\
\hline
    \end{tabular}
\end{table}

The top two rows of Table \ref{t3tab:results} present a comparison of the different modeling approaches when propensity scores are included. Specifically, the results from the models with $d=4$ are presented for each modelling approach, as these have been chosen best fitting models according to DIC. In contrast to the na\"ive analysis in Web Table 1 that shows a negative effect of PEG, analyses under both Assumption 3 and Assumption 4 show that there is a significant, positive SACE of PEG surgery on the outcome of BMI at 18 months post-baseline. Holding all else constant and assuming PEG insertion at or immediately after baseline, BMI at 18 months increases by about 3 units for those individuals who receive treatment when compared to those who are not treated among survivors. The effect of time to treatment is also positive but not statistically significant. Noting that a higher time to PEG ($T_Z$) leads to a shorter time from PEG to the time of outcome measurement, these results seem to suggest that the positive effect of PEG on BMI decreases gradually after the treatment, i.e., the possibility for a waning long-term effect. Additionally, the approach assuming MCAR indicates that the estimated effects of time to treatment and binary treatment are smaller in magnitude than those estimated from the two approaches under Assumptions 3 and 4, respectively.

The lower half of Table \ref{t3tab:results} presents the results from the three approaches without adjustment for propensity scores in the estimation of the SACE of PEG treatment, showing some changes in the magnitude of the effect estimates.  On the one hand, removing the propensity score from the models in Approaches 1 and 2, the time to treatment effect remains positive and is now significant. On the other hand, in these same models, the coefficient estimate for the binary treatment indicator is drastically smaller in magnitude and no longer statistically significant. In the model that assumes MCAR and does not use propensity scores, the coefficient estimate for the binary treatment indicator changes direction and is non-significant.

In addition, the results, not provided, are not sensitive to the assumption of monotonicity. This may be due to the small proportion of individuals in $DL$ strata when all four strata are considered. When monotonicity is not assumed, most patients are in the $LL$ and $DD$ strata, with a $LD$ and $DL$ strata comprising less than 20\% of the individuals in total. It is conceivable then that reallocating such a small proportion of individuals when removing the $DL$ stratum would not substantially change the effect estimates of the other strata.

Overall, the results in our analysis demonstrate a positive SACE of PEG, in contrast to inconclusive results in magnitude, direction, and significance from previous observational studies of PEG in association with weight or BMI. Given that our methodology is developed to account for the many complicating issues of an observational study that were not addressed in previous studies, it is likely that the positive effect of PEG estimated in this analysis is more definitive. As data collection is still ongoing in the ALS Clinic Registry, the proposed approach can be applied to this registry in the future, potentially further validating our current findings.

\section{Discussion}
\label{sec:dicussion}

Our current work represents the very first attempt to leverage the rich data in the Emory ALS Clinic Registry for assessing the effect of PEG in a large ALS population. Our proposed framework for causal inference addresses several challenges in the analysis of this data that have not been accounted for in previous studies of PEG. Our results show a positive effect of PEG placement on survivors that may be waning over time after insertion. Our approach can be directly applied or extended to evaluate effects of PEG or other palliative procedures such as the non-invasive positive pressure ventilation (NIPPV) \citep{miller2009practice} on additional outcomes  that are of interest to clinicians including quality of life and disease progression.

One limitation of our analysis is that we do not have data on the usage of PEG tube after its insertion. Of note, patients with PEG may elect not to use PEG for nutrition delivery from time to time, thus the level of usage, similar to treatment adherence, could have an impact on patient outcomes such as BMI. Additionally, while the use of propensity scores within the principal stratification framework allows for the estimation of an unbiased principal effect using observational data, the reliability of removing bias via propensity score adjustment is predicated on the assumption of strongly ignorable treatment assignment, which means there must be no unmeasured confounders. Partly based on our experience from this work, a future prospective observational study, currently in the design and planning stage, will collect usage data for palliative procedures such as PEG and NIPPV as well as additional potential confounders that are not collected or validated in standard clinical settings. In addition, future applications of the proposed methods to other data with richer measurements of confounders should further demonstrate the reduction of confounding.

Making adjustments for missing outcome data in the context of causal inference typically requires strong assumptions about the ignorability of the missing data mechanism and creativity in the modeling framework. In our data application, the similar results under Assumptions 3 and 4 indicate that the findings are not sensitive to the latent ignorable assumption for this data, though we cannot empirically test this assumption. When in doubt, further stratification by missingness of outcome data and within the principal stratification framework may allow for additional flexibility in the assumptions imposed on the missing data mechanism.

Future consideration may also be given to jointly modeling the propensity score with the outcome model, missing data model, and principal strata model in a Bayesian framework, similar in spirit with \citet{zigleretal2013} and \cite{zigler2014uncertainty}. This would allow the quantities observed by in each of these three models to affect the posterior of propensity score in each MCMC iteration. While this could provide a more robust propensity score adjustment, \citet{zigleretal2013} showed that the feedback between model stages in joint modeling can cause biased causal effect estimates if individual covariates are not also adjusted for in the outcome model. This bias should be accounted for if one uses joint modeling of the four models of outcome, missing data mechanism, propensity scores, and principal strata.

%\nocite{*}
\bibliographystyle{apa}
\bibliography{paper3}

\end{document}